\documentclass[twocolumn,prl,preprintnumbers,amsmath,amssymb,nofootinbib]{revtex4-2} 
\usepackage{slashed}
\usepackage{mathtools}
\usepackage{amsfonts}
\usepackage{amssymb}
\usepackage{epsfig}
\usepackage{rotating}
\usepackage{url}
\usepackage{times}
\usepackage{color}
\usepackage{bm}
\usepackage{xcolor,colortbl}
\usepackage{hyperref}

\newcommand{\GeV}{\ensuremath{{\rm GeV}}}
\newcommand{\meV}{\ensuremath{{\rm meV}}}
\newcommand{\mum}{\ensuremath{\mu {\rm m}}}
\newcommand{\nm}{\ensuremath{{\rm nm}}}
\newcommand{\mm}{\ensuremath{{\rm mm}}}
\newcommand{\Mpc}{\ensuremath{{\rm Mpc}}}
\newcommand{\mssq}{\ensuremath{{\rm m/s^2}}}
\newcommand{\mass}{\ensuremath{{\cal M}}}

\newcommand{\nablab}{\bar{\nabla}} 
\newcommand{\grad}{\ensuremath{\vec{\nabla}}}

\newcommand{\rhob}{{\bar{\rho}}}
\newcommand{\Pb}{{\bar{P}}}
\newcommand{\Ealp}{\ensuremath{E}}
\newcommand{\phib}{{\bar{\phi}}}

\newcommand{\RMOND}{RMOND~}
\newcommand{\lambdas}{\lambda_s}

\newcommand{\PhiE}{\hat{\Phi}}

\newcommand{\Kcal}{{\cal K}}

\newcommand{\phit}{\tilde{\phi}}
\newcommand{\ggc}{g_{{\rm GGC}}}
\newcommand{\MC}{M_{{\rm GGC}}}
\newcommand{\Fh}{F}
\newcommand{\Ah}{{A}}
\newcommand{\Jh}{{J}}
\newcommand{\AC}{\hat{A}}
\newcommand{\FC}{\hat{F}}

\newcommand{\Mpt}{\tilde{M}_p}
\newcommand{\YHe}{Y_{\rm He}}

\newcommand{\Dcal}{{\cal D}}
\newcommand{\Mcal}{{\cal M}}

\newcommand{\Fcal}{{\cal F}}
\newcommand{\Xcal}{{\cal X}}
\newcommand{\Qcalb}{\bar{{\cal Q}}}
\newcommand{\Zcal}{{\cal Z}}
\newcommand{\Qcal}{{\cal Q}}
\newcommand{\Ycal}{{\cal Y}}

\newcommand{\Xcalb}{\bar{{\cal X}}}

\newcommand{\Jcal}{{\cal J}}

\newcommand{\metM}{\ensuremath{g}}

\newcommand{\Gqs}{\ensuremath{\hat{G}}}
\newcommand{\Gt}{\ensuremath{\tilde{G}}}
\newcommand{\GN}{\ensuremath{G_{N}}}
\newcommand{\KB}{\ensuremath{K_{B}}}
\newcommand{\cad}{c_{\rm ad}}

\DeclareMathSizes{10}{10}{7}{6}

\definecolor{orange}{rgb}{1,0.5,0}
\definecolor{darkorange}{rgb}{0.69,0.33,0.13}
\definecolor{fidcol}{rgb}{0.7,0,0}

\begin{document}
	
\title{New Relativistic Theory for Modified Newtonian Dynamics}

\author{Constantinos Skordis}
\email{skordis@fzu.cz}
\author{Tom~Z\l o\'{s}nik}
\email{zlosnik@fzu.cz}

\affiliation{CEICO, Institute of Physics (FZU) of the Czech Academy of Sciences, Na Slovance 1999/2, 182\,\,21, Prague, Czech Republic}

\begin{abstract}
We propose a relativistic gravitational theory leading to modified Newtonian dynamics, a paradigm that explains
the observed universal galactic acceleration scale and related phenomenology.
We discuss phenomenological requirements leading to its construction and demonstrate its agreement
with the observed cosmic microwave background and matter power spectra on linear cosmological scales.
We show that its action expanded to second order is free of ghost instabilities and discuss its possible embedding in a more fundamental theory. 
\end{abstract}

\maketitle

\paragraph{Introduction. --}
Alternative theories of gravity to general relativity (GR) have received immense interest in the past 20 years 
or so~\cite{Jain:2010ka,Clifton:2011jh}. The driving force behind this interest is not so much that gravity has not 
been tested  in a large region of parameter space~\cite{BakerPsaltisSkordis2014}, but, more importantly, 
 the cosmological systems residing in some parts of that region exhibit behavior from which dark matter (DM) and dark energy (DE), collectively called the dark sector, 
are inferred.

While most investigations deal with DE,
the hypothesis that the DM phenomenon is due to gravitational degrees of freedom (d.o.f.) has received less attention~\cite{Cembranos:2008gj,Chamseddine:2013kea,Arroja:2015yvd,Sebastiani:2016ras,Casalino:2018tcd,Bettoni:2013zma,Mendoza:2012hu,Rinaldi:2016oqp,Koutsoumbas:2017fxp,Diez-Tejedor:2018fue,Milgrom:2018bit}.
Earliest evidence for the existence of DM~\cite{Oort1932,Zwicky:1933gu,Smith:1936mlg} 
was later supported by observations of the motion of stars within galaxies \cite{Rubin:1970zza,RubinEtAl1980}.
 Milgrom proposed~\cite{Milgrom1983a,Milgrom1983b,Milgrom1983c} that this could, instead, result from modifying the inertia or dynamics of baryons
or  the gravitational law  at accelerations smaller than $a_0\sim 1.2\times 10^{-10} \mssq$.
The latter is further explored in~\cite{BekensteinMilgrom1984} where
if gradients of the potential $\Phi$ are smaller than $a_0$,
nonrelativistic gravity is effectively governed by
\begin{equation}
 \grad \cdot\left(  \frac{|\grad\Phi|}{a_0} \grad \Phi \right)  = 4 \pi \GN \rho.
\label{eq_AQUAL}
\end{equation}
Here, $\GN$ is the Newtonian gravitational constant, and $\rho$ the matter density.
These models are referred to as modified Newtonian dynamics (MOND). 

Much work has gone into deducing astrophysical consequences of MOND, 
its consistency with data \cite{Milgrom1988,Kent1988,GerhardSpergel1992,Begeman:1991iy,Milgrom:1995hz,Sanders:1996ua,Sanders1999,McGaugh:2000sr,SandersMcGaugh2002,McGaugh:2005qe,BinneyFamaey2005,BekensteinMagueijo2006,GentileEtAl2007,MagueijoMozaffari2011,Famaey:2011kh,Heesetal2015,MargalitShaviv2015,McGaughEtAl2016,HodsonZhao2017,BilekEtAl2017,Sanders:2018mnk,LelliEtAl2019,Petersen:2020vks},
and alternative DM based explanations of this law~\cite{Blanchet:2006yt,Blanchet:2009zu,Berezhiani:2015pia,Berezhiani:2015bqa} 
It is inherently nonrelativistic and, thus, difficult to test in cosmological settings (but see \cite{PardoSpergel2020}) as systems such as the cosmic microwave background (CMB) require a relativistic treatment.
CMB physics involves only linearly perturbing a Friedmann-Lema\^itre-Robertson-Walker (FLRW) background,
 making it a particularly useful system, devoid of nonlinear modeling systematics,
 for testing relativistic MOND (RMOND). 
Relativistic theories that yield MOND behavior have been proposed
\cite{BekensteinMilgrom1984,Bekenstein1988,Sanders1997,Bekenstein2004,NavarroVanAcoleyen2005,ZlosnikFerreiraStarkman2006,Sanders2007,Milgrom2009,BabichevDeffayetEsposito-Farese2011,DeffayetEsposito-FareseWoodard2011,Woodard2014,Khoury2014,Blanchet:2015sra,Hossenfelder2017,Burrage:2018zuj,Milgrom:2019rtd,DAmbrosio:2020nev} 
 making clear predictions regarding gravitational lensing and cosmology.
In cases where the CMB and matter power spectra (MPS) have been computed, no theory has been shown to fit all of the cosmological data while
preserving MOND phenomenology in galaxies~\cite{SkordisEtAl2005,BourliotEtAl2006,DodelsonLiguori2006,ZuntzEtAl2010,XuEtAl2014,DaiStojkovic2017,DaiStojkovic22017,ZlosnikSkordis2017,TanWoodard2018}, (though see \cite{Sanders2005}).

We present the first \RMOND theory which reproduces galactic and lensing phenomenology 
similar to the Bekenstein-Sanders Tensor-Vector-Scalar (TeVeS) theory \cite{Sanders1997,Bekenstein2004} and, unlike TeVeS,
 successfully reproduces the key cosmological observables: CMB and MPS. We describe its construction, discuss its cosmology and show that 
it is devoid of ghost instabilities. We discuss open questions
and possibilities toward its more fundamental grounding.

\paragraph{Phenomenological requirements. --} 
\RMOND theories have always been constructed on phenomenological grounds rather than based 
on fundamental principles. Quite likely the reason is that the MOND law is empirical, and even the observation that 
it is scale invariant~\cite{Milgrom1997,Milgrom2008} has not yet led to a definitive conclusion as to how this invariance 
could lead to a MOND gravitational theory. \RMOND theories should obey the principle of general covariance 
and the Einstein equivalence principle. These are, however, 
do not provide any guidance as to how \RMOND  should look like. Indeed, many theories obeying these have 
nothing to do with MOND, and many \RMOND theories obeying these same principles are in conflict with observations.
Principle-based MOND theories include \cite{Milgrom:1998sy,Klinkhamer:2011un,Verlinde2016}, however, these are nonrelativistic. 
Still, the phenomenological approach, that we also follow, can provide valuable guidance toward a more fundamental theory.

What are the necessary phenomenological facts that any successful MOND theory should lead to? 
It must (i) return to GR (hence, Newtonian gravity) when $\grad\Phi\gg a_0$ in quasistatic situations while (ii)
 reproducing the MOND law \eqref{eq_AQUAL} when $\grad\Phi \ll a_0$.
It should also (iii) be in harmony with cosmological observations including the CMB and MPS,
(iv) reproduce the observed gravitational lensing of isolated objects without DM halos,
and (v) propagate tensor mode gravitational waves (GWs) at the speed of light.

We consider each requirement in turn. 
 Clearly, (i) means that when $|\grad \Phi| \gg a_0$, the standard Poisson equation $\grad^2\Phi = 4\pi \GN \rho$ holds while (ii) means that
when $|\grad \Phi| \ll a_0$ the MOND equation \eqref{eq_AQUAL} holds.
While in  many cases~\cite{DeffayetEsposito-FareseWoodard2011,Woodard2014,ZlosnikFerreiraStarkman2006}
 the transition between (i) and (ii) depends only on $|\grad \Phi|$, in TeVeS it is facilitated by
a scalar d.o.f. $\varphi$. We follow the latter and assume that the physics encapsulated by (i) and (ii) fits within the TeVeS framework.

 A template nonrelativistic action then, is
\begin{equation}
S =  \int d^4x  \left\{
  \frac{1}{8\pi \Gqs} \left[ |\grad\PhiE|^2 
+ \Jcal(\Ycal)
\right]
+  \Phi \rho\right\},
\label{scalar_AQUAL_action}
\end{equation}
where $\Phi = \PhiE + \varphi$ is the potential that couples universally to matter, $\Gqs$ is a constant and $\Ycal = |\grad\varphi|^2$.
The field  $\varphi$ obeys \; \;$\grad \cdot[ (d\Jcal/d\Ycal)\grad\varphi] = 4 \pi \Gqs \rho$ while  $\PhiE$ obeys the Poisson equation  $\grad^2 \PhiE = 4 \pi \Gqs \rho$.
Emergence of MOND is then ensured if $\Jcal \rightarrow \frac{2 \lambdas}{3 (1 + \lambdas) a_0} \Ycal^{3/2}$ as $\grad\varphi\rightarrow 0$. It is in this limit that $a_0$ appears.

For a point source of mass $M$, the MOND-to-Newton transition occurs at $r_M \sim  \sqrt{(\GN M/a_0)}$.
A MOND force $\sim \sqrt{\GN M a_0}/r$ lends its way trivially to a Newtonian force $\GN M/r^2$  as $r\ll r_M$ but in the inner Solar System
this is not sufficient. Corrections to $r^{-2}$ due to $\varphi$ will compete with  the post-Newtonian force  $\sim (\GN M)^2/r^3$, and these are constrained
at Mercury's orbit to less than $\sim10^{-4}$~\cite{Will2014,Will2018}.  
Suppressing these may happen either through screening or tracking. In the former,  $\varphi$ is screened 
at large $\grad\varphi$  so that $\Phi \approx \PhiE$ while in the latter $\varphi \rightarrow \PhiE/\lambdas$, so that $\GN = (1+ 1/ \lambdas)\Gqs $. We model both with $\lambdas$  since screening is equivalent to $\lambdas\rightarrow \infty$.
In terms of $\Jcal$, tracking happens if $\Jcal \rightarrow \lambdas \Ycal$, 
while screening occurs if $\Jcal$ has terms $\Ycal^p$ with $p\ge 3/2$ (this may be in conflict with Mercury's orbit even as $p\rightarrow \infty$) or via higher-derivative terms absent from \eqref{scalar_AQUAL_action}.

Consider requirement (iii), that is, successful  cosmology. 
In \eqref{scalar_AQUAL_action} we have a new d.o.f. $\varphi(\vec{x})$ and we expect that the same will appear in cosmology,
albeit with a time dependence, i.e. $\phib(t)$.
Consider a flat FLRW metric so that  $g_{00} = -N^2$ and $g_{ij} = a^2 \gamma_{ij}$  where $N(t)$ is the lapse function and $a(t)$ 
the scale factor. What should the expectation for
a cosmological evolution of $\phib(t)$ be? The MOND law for galaxies is silent regarding this matter. There is, however, another empirical 
law which concerns cosmology:
the existence of sizable amounts of energy density scaling precisely as $a^{-3}$. Within the DM paradigm such a law is a natural consequence of 
particles obeying the collisionless Boltzmann equation. The validity of this law has been tested~\cite{KoppEtAl2018,IlicEtAl2020} and 
during the time between radiation-matter equality and recombination it is valid within
an accuracy of $\sim 10^{-3}$. Do scalar field models leading to energy density scaling as $\rhob \sim a^{-3}$ exist?

 The answer is yes: shift symmetric $k$ essence.
 It has been shown~\cite{Scherrer2004} that a scalar field with Lagrangian $\sim \Kcal(\Xcalb)$ where $\Xcalb = \dot{\phib}^2/N^2$, 
leads to dust (i.e.  $\rhob \sim a^{-3}$) plus cosmological constant (CC) solutions provided $\Kcal(\Xcalb)$ has a minimum at $\Xcalb = \Xcal_0 \ne 0$. 
Such a model is the low energy limit of ghost condensation~\cite{ArkaniHamedEtAl2003,ArkaniHamed:2005gu} although the latter also
contains higher derivative terms $\sim (\square\phi)^2$ in its action.  The FLRW action is
\begin{align}
S =&  
\frac{1}{8\pi \Gt} \int d^4x N a^3 \left[
 -\frac{3H^2}{N^2}
+ \Kcal(\Qcalb) \right] + S_m[g]
\label{sculpted_FRW_action}
\end{align}
where 
$\Qcalb = \dot{\phib}/N$
and $H = \dot{a}/a$.
Interestingly,
 \eqref{scalar_AQUAL_action} and 
\eqref{sculpted_FRW_action}  are shift symmetric in $\varphi$ and $\phib$ respectively. 

We propose that the MOND analog on FLRW is given by
\eqref{sculpted_FRW_action} with 
\begin{equation}
\Kcal = -2\Lambda + \Kcal_2(\Qcalb - \Qcal_0)^2 + \ldots
\label{Kcal_expansion}
\end{equation}
 where $\Lambda$ is the CC,  $\Kcal_2$ and $\Qcal_0$  parameters and $(\ldots)$ denote higher powers in this expansion.
Expanding in $\Qcal-\Qcal_0$ rather than $\Xcal-\Xcal_0$ is the most general expansion leading to dust solutions and includes the $\Kcal(\Xcalb)$ case.
The CC in this model remains a freely specifiable parameter, just as in the $\Lambda$-cold dark matter ($\Lambda$CDM) model. Following~\cite{ArkaniHamedEtAl2003,ArkaniHamed:2005gu},
we call this the (gravitational) Higgs phase.

Requirement (iv), that is, correct gravitational lensing without DM, requires a relativistic theory. 
A minimal theory for \RMOND is a scalar-tensor theory\cite{BekensteinMilgrom1984} with the scalar providing for a conformal factor between two metrics.
However, since null geodesics are unaltered by conformal transformations, such theories cannot produce enough lensing from baryons in the MOND regime.
Sanders solved the lensing problem by changing the conformal into a disformal transformation~\cite{Sanders1997} using a unit-timelike vector field, 
 incorporated by Bekenstein~\cite{Bekenstein2004} into TeVeS. The unit-timelike vector 
has component $A^0 \sim \sqrt{-g^{00}}$ and this ensures that the two metric potentials are equal (as in GR), so that solutions which mimic DM
also produce the correct light deflection.

Meanwhile the anisotropic scaling of the MOND law $\sim |\grad\varphi|^3$ compared with 
a well-behaved cosmology implying terms like $\dot{\phib}^2$ and $\dot{\phib}^4$, heuristically implies (gravitational) Lorentz violation.
A good way of introducing such an ingredient is via a unit-timelike vector field $A_\mu$,  much like the spirit of the Einstein-{\AE}ther theory~\cite{Dirac1962,JacobsonMattingly2000},
and TeVeS~\cite{Sanders1997,Bekenstein2004}.

The advanced Laser Interferometer Gravitational Observatory (LIGO) and
Virgo interferometers \cite{TheLIGOScientific:2017qsa} observed GWs from a binary neutron star merger.
Combined with electromagnetic observations~\cite{Goldstein:2017mmi,Savchenko:2017ffs}, this strongly constrains the GW tensor mode speed to be effectively equal to that of light. 
By analyzing the tensor mode speed, TeVeS has been shown~\cite{BoranEtAl2017,GongEtAl2018,HouGong2018,SkordisZlosnik2019}  
 to be incompatible with the LIGO-Virgo observations for any choice of parameters. 
The necessary d.o.f. $\phi$ and $A_\mu$ are also ingredients of TeVeS,
only there, a second metric was introduced as a combination of $g_{\mu\nu}$, $\phi$ and $A_\mu$. In~\cite{ZlosnikFerreiraStarkman2006b},
$\phi$ and $A_\mu$ were combined into a timelike (but not unit) vector $B_\mu$, and it was shown that
 TeVeS may be equivalently formulated with a single metric  $g_{\mu\nu}$  minimally coupled to matter, 
and  $B_{\mu}$ with a noncanonical and rather complicated kinetic term.
A general class of theories based on the pair $\{g_{\mu\nu},B_\mu\}$ was uncovered~\cite{SkordisZlosnik2019} where the tensor mode speed equals the speed of light in all situations, satisfying requirement (v).

\paragraph{The new theory. --}
A subset of the general class~\cite{SkordisZlosnik2019}  depends on a scalar $\phi$ and
unit-timelike vector $A^\mu$ such that~\cite{SkordisZlosnikInPreparation}
\begin{align}
S =&  \int d^4x \frac{\sqrt{-\metM}}{16\pi \Gt} \bigg[ R  
 - \frac{\KB}{2}  \Fh^{\mu\nu} \Fh_{\mu\nu} 
+ 2  (2-\KB) \Jh^\mu\nabla_\mu \phi
\nonumber
\\
&
- (2-\KB) \Ycal
- \Fcal(\Ycal,\Qcal)
 - \lambda(\Ah^\mu \Ah_\mu+1)\bigg] + S_m[g]
\label{NT_A_action}
\end{align}
where $\Fh_{\mu\nu} =2\nabla_{[\mu} \Ah_{\nu]}$, $\Jh_\mu = \Ah^\alpha \nabla_\alpha \Ah_\mu$,
and the Lagrange multiplier $\lambda$ imposes the unit-timelike constraint on $\Ah_\mu$. In addition $\Fcal(\Ycal,\Qcal)$ is a free function of 
$\Qcal = \Ah^\mu \nabla_\mu \phi$ and $\Ycal= q^{\mu\nu} \nabla_\mu \phi \nabla_\nu \phi$ where 
$q_{\mu\nu} = g_{\mu\nu} + \Ah_\mu \Ah_\nu$ is the three-metric orthogonal to $\Ah^\mu$.
Notice that \eqref{NT_A_action} is shift symmetric under $\phi \rightarrow \phi + \phi_0$.

On FLRW $\phi= \phib(t)$ while $\Ah_0 = -N$ and $\Ah_i = 0$, hence $\Ycal =0$ and $\Qcal = \Qcalb$. We define
$\Kcal(\Qcalb) = -\frac{1}{2}\Fcal(0,\Qcalb)$ so that \eqref{NT_A_action} turns  precisely into  \eqref{sculpted_FRW_action}, which we have 
argued that it satisfies requirement (iii).

In the weak-field quasistatic limit, we set $g_{00} = -1-2\Psi$ and $g_{ij} = (1-2\Phi)\gamma_{ij}$ and
  assume that $\Ah^\mu$ aligns with the time direction so that $\Ah^0 = 1-\Psi$ and $\Ah^i=0$.
 The scalar is expanded as $\phi = \phib + \varphi$ with $\dot{\varphi} \ll |\grad\varphi|$ and $\dot{\phib}$ may be set to its (late Universe)
FLRW minimum $\Qcal_0$. Hence, $\Qcal =  (1-\Psi)  \Qcal_0 $.
 Then \eqref{NT_A_action}  leads to $\Psi = \Phi$ which can be subbed back to get
\begin{align}
S =& - \int d^4x  \bigg\{
\frac{2-\KB}{16\pi \Gt} \bigg[ |\grad\Phi|^2 
- 2  \grad \Phi \grad \varphi
\nonumber
\\
&
+ |\grad \varphi|^2
- \mu^2 \Phi^2
+ \Jcal(\Ycal)
\bigg]
+  \Phi \rho\Bigg\}
\label{NT_quasi_Phi}
\end{align}
where $\Jcal(\Ycal) =  \Fcal(\Ycal,\Qcal_0)/ (2-\KB)$.
Compared with \eqref{scalar_AQUAL_action} a new term appears 
which looks like a ``mass term''  for  $\Phi$, with $\mu = \sqrt{\frac{2 \Kcal_2}{2-\KB}} \Qcal_0 $.
The solution for $\Phi$ will be as obtained from  \eqref{scalar_AQUAL_action} only for $r \lesssim r_C$ where $r_C \sim \left(r_M \mu^{-2}\right)^{1/3}$, and oscillatory for $r \gtrsim r_C$.
We require $\mu^{-1} \gtrsim 1 \Mpc$ so that MOND behavior according to \eqref{scalar_AQUAL_action} may still be attained in galaxies.
Thus, the quasistatic limit has at least three parameters: $\lambdas$, $a_0$ and $\mu$.

While matter couples only to $\Phi$, gravity comes with 
two potentials $\Phi$ and $\varphi$ whose action is not diagonal but contains the  mixing term $J^\mu\nabla_\mu\phi \rightarrow  \grad\Psi \cdot \grad\varphi$.
Without the latter, $\varphi$ decouples and  no  modification of gravity arises in this situation, apart from $\mu^2\Phi$ which 
   is akin to ghost condensation~\cite{ArkaniHamedEtAl2003,ArkaniHamed:2005gu}.
Diagonalizing by setting $\Phi = \PhiE + \varphi$ and identifying $\Gt = (1 - \frac{\KB}{2})\Gqs$ turns  \eqref{NT_quasi_Phi} into \eqref{scalar_AQUAL_action} (plus the $\mu^2 \Phi^2$ term).
Since, $\Psi =\Phi$, \eqref{NT_quasi_Phi} leads to the right lensing whenever the solution for $\Phi$ mimics DM.
 This satisfies requirements (i), (ii) and (iv).

\paragraph{Cosmological observables. --}
The theory just presented was constructed to lead to a FLRW universe resembling $\Lambda$CDM. Given a general $\Kcal(\Qcal)$,
we define the energy density as $8\pi \Gt \rhob =  \Qcal \frac{d\Kcal}{d\Qcal} - \Kcal$ and pressure as $8\pi \Gt\Pb = \Kcal$ so that the usual FLRW equations are satisfied.
The field equation for $\phib$ may be integrated once to give $\frac{d\Kcal}{d\Qcal} = \frac{I_0}{a^3}$ for initial condition $I_0$.
When $\Kcal$ obeys the expansion \eqref{Kcal_expansion}, then $\Qcal = \Qcal_0 + I_0/a^3 + \ldots$, so that
$\rhob = \rhob_0/a^3 + \ldots$, where $8\pi \Gt \rhob_0 = \Qcal_0 I_0$.
 The pressure is $\Pb =  w_0 \rhob_0/a^6  + \ldots$ 
where $w_0=  \frac{8\pi \Gt\rhob_0}{4\Qcal_0^2 \Kcal_2}$ is the equation of state at $a=1$, that is,
 $w = w_0/a^3 + \ldots$
so that $\Pb = w \rhob$. A time-varying $w$ implies an adiabatic sound speed 
$\cad^2 = d\Pb/d\rhob = \frac{d\Kcal/d\Qcal }{   \Qcal \; d^2\Kcal/d\Qcal^2 } $ and
if $\Kcal$ obeys  \eqref{Kcal_expansion} then $\cad^2 = 2 w_0/a^3 + \ldots$.
Clearly, $w\ge 0$ and  $\cad^2\ge 0$, where the zero point is reached as $a\rightarrow \infty$.
As the solution depends on the initial condition $I_0$,  the density $\rhob$ is not (classically) predicted.

For a proper  cosmological matter era in the Higgs phase we need $w_0$ to be sufficiently small.
Observations~\cite{KoppEtAl2018,IlicEtAl2020} give $w\lesssim 0.02$ at $a\sim 10^{-4}$,
hence, $w_0 \lesssim  2 \times 10^{-14}$.  Meanwhile,  $\mu^{-1} \gtrsim \Mpc$ in order not to spoil the MOND behavior, leading to $w_0>\frac{3H_0^2 \Mpc^2 \Omega_0}{2(2-\KB)}\gtrsim 10^{-8}$. Unless 
the effect of the $\mu$ term in \eqref{NT_quasi_Phi}  is alleviated in some future theory,
the Higgs phase cannot be extended too long in the past, and higher terms in  \eqref{Kcal_expansion} must be taken into consideration. 
Within the present setup, one can arrange this with a function $\Kcal(\Qcal)$ which 
suppresses $w$ and $\cad^2$ during most of the cosmic evolution. Examples are $\Kcal = 2 \Kcal_2 \Zcal_0^2 \left[ \cosh(\Zcal)-1 \right]$ 
 (``Cosh function'') and $\Kcal = 2 \Kcal_2 \Zcal_0^2 \left[ e^{\Zcal^2} -1\right]$ (``Exp function'') where $\Zcal = (\Qcal-\Qcal_0)/\Zcal_0$.

The tight coupling of baryons to photons in the early Universe leads to Silk damping 
and wipes out all small-scale structure in baryons, preventing the formation of galaxies in the late Universe.
Within GR, cold DM sustains the gravitational potentials during the tight coupling period,  driving the formation of galaxies
and affecting the relative peak heights of the CMB as further corroborated by e.g. the Planck satellite~\cite{Aghanim:2018eyx}. Checking whether 
this theory fits the CMB and MPS spectra requires studying linear fluctuations on FLRW.

We consider scalar modes in the Newtonian gauge  so that $g_{00} =  -(1 + 2 \Psi)$, $g_{0i}=0$ and $g_{ij} = a^2  (1 - 2\Phi) \gamma_{ij}$
and perturb the scalar as $\phi = \phib + \varphi$ and the vector as $\Ah_\mu = \{ -1 - \Psi, \,  \grad_i \alpha\}$. 
The perturbed Einstein, vector and scalar equations, then depend on the new scalar modes 
$\varphi$ and $\alpha$ and their derivatives. The shear equation remains as in GR, as do the usual perturbed Boltzmann equations for 
 baryon, photons and neutrinos, since they couple only to $g_{\mu\nu}$.

Setting $\chi \equiv \varphi +   \dot{\phib} \alpha$, $\gamma \equiv \dot{\varphi} - \dot{\phib}  \Psi$,  $\Ealp \equiv \dot{\alpha} + \Psi$
and defining the  density contrast $\delta$ and momentum divergence $\theta$  via
\begin{align}
\delta \equiv&  \frac{1+ w}{\dot{\phib} \cad^2} \gamma 
+ \frac{1}{8\pi \Gt a^2 \rhob} \grad^2 \left[ \KB \Ealp + \left(2- \KB \right) \chi \right]
\label{delta_field_relation}
\\
\theta \equiv& \frac{\varphi}{\dot{\phib}}
\label{theta_field_relation}
\end{align}
the Einstein equations take the same form as in GR, i.e. $\delta G^0_{\;\;0} = 8\pi G \sum_I \rhob_I \delta_I$ and
 $\delta G^0_{\;\;j} =  - 8\pi G \sum_I (\rhob_I +\Pb_I) \grad_j \theta_I $   where the index $I$ runs over all
matter species including the new variables $\delta$ and $\theta$. 
These obey standard fluid equations  
\begin{align}
 \dot{\delta} =& 3  H \left(w \delta - \Pi\right)     + ( 1+ w) \left( 3   \dot{\Phi} - \frac{k^2}{a^2} \theta  \right)
\label{delta_phi_dot_cv0}
\\
 \dot{\theta} =& 3\cad^2  H  \theta +  \frac{\Pi}{1+w} +   \Psi 
\label{theta_phi_dot}
\end{align}
but with nonstandard pressure contrast:
\begin{equation}
\Pi = \cad^2 \delta  - \frac{\cad^2}{8\pi \Gt a^2 \rhob} \grad^2 \left[ \KB \Ealp + \left(2- \KB \right) \chi \right]
\label{Pi_delta_E_alpha}
\end{equation}
\begin{figure}[t]
\begin{center}
\includegraphics[width=8.6cm]{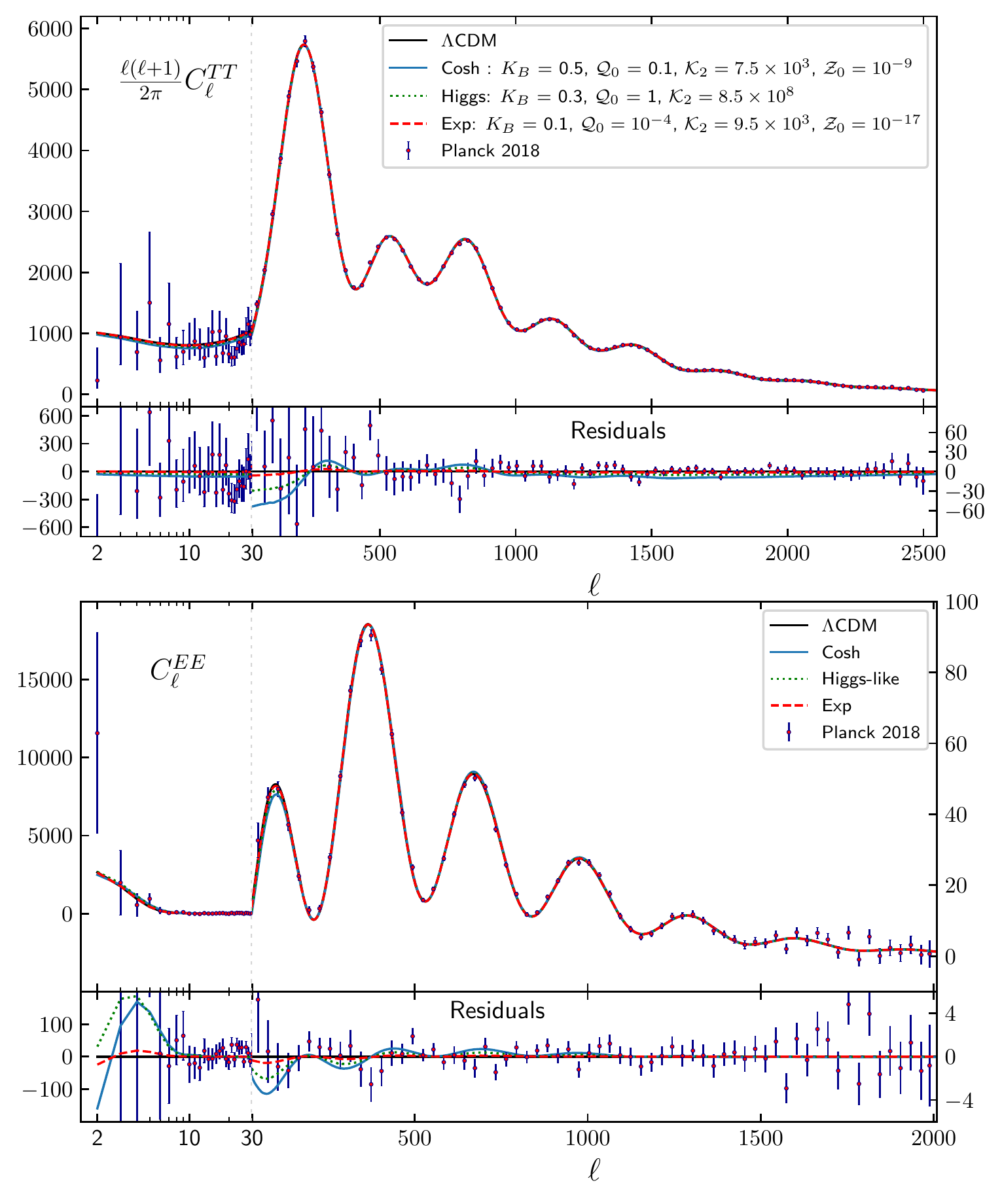}
\end{center}
\vspace{-7mm}
\caption{The CMB   temperature (T) $C_\ell^{TT}$ and $E$-mode polarization  $C_\ell^{EE}$ 
angular power spectra  for $\Lambda$CDM  
and this theory for a collection of functions and parameter values.
The $\Lambda$CDM parameters are angular acoustic scale $100\theta_s = 1.04171$, DM density $\Omega_c h^2 = 0.1202$, baryon density $\Omega_b h^2 = 0.02235$,
reionization optical depth $\tau = 0.049$, helium fraction $\YHe = 0.242$, primordial scalar amplitude $10^9 A_s = 2.078$ and spectral index $n_s = 0.963$,
 while the  MOND curves deviate from these
 within  $\sim \{0.07,0.33,3.98,14.29,1.57,0.58,2.60\}$ percent.
MOND models have $\lambdas = \infty$ and 
their other parameters are shown in the $C_\ell^{TT}$ panel, with $\Qcal_0$ and $\Zcal_0$ in $\Mpc^{-1}$.
The ``Higgs-like'' function parameters are incompatible with a MOND limit.
}
\vspace{-6mm}
\label{Fig:cl}
\end{figure}
Hence, the resulting system is not equivalent to a dark fluid: the nonstandard pressure, thus defined, does not close under the fluid variables but, rather,
depends on the vector field perturbations $\alpha$ and $\Ealp$. The latter evolves with
\begin{align}
  \KB \left(\dot{\Ealp} + H \Ealp \right)
=&  \frac{d\Kcal}{d\Qcal} \chi
- (2 - \KB) \bigg[ \frac{ \dot{\phib} }{ 1+ w } \Pi
\nonumber
\\
&
+ \left(H + \dot{\phib} \right) \chi
-3\cad^2 H \dot{\phib}  \alpha
\bigg]
\end{align}

Cosmologically, the necessary additional free parameters to $\Lambda$CDM are $\lambdas$ (influencing the effective cosmological gravitational strength),
 $\KB$, $\Kcal_2$ (or equivalently $w_0$) and $\Qcal_0$. These fix $\mu$ appearing in the quasistatic regime.
More elaborate functions $\Kcal(\Qcal)$ introduce further parameters, e.g. $\Zcal_0$ in the case of the ``Cosh'' or ``Exp'' functions above.
Note that $a_0$ does not appear in the linear cosmological regime but will play a role once nonlinear terms from $\Fcal(\Ycal,\Qcal)$ kick in.

\begin{figure}[t]
\begin{center}
\includegraphics[width=8.6cm]{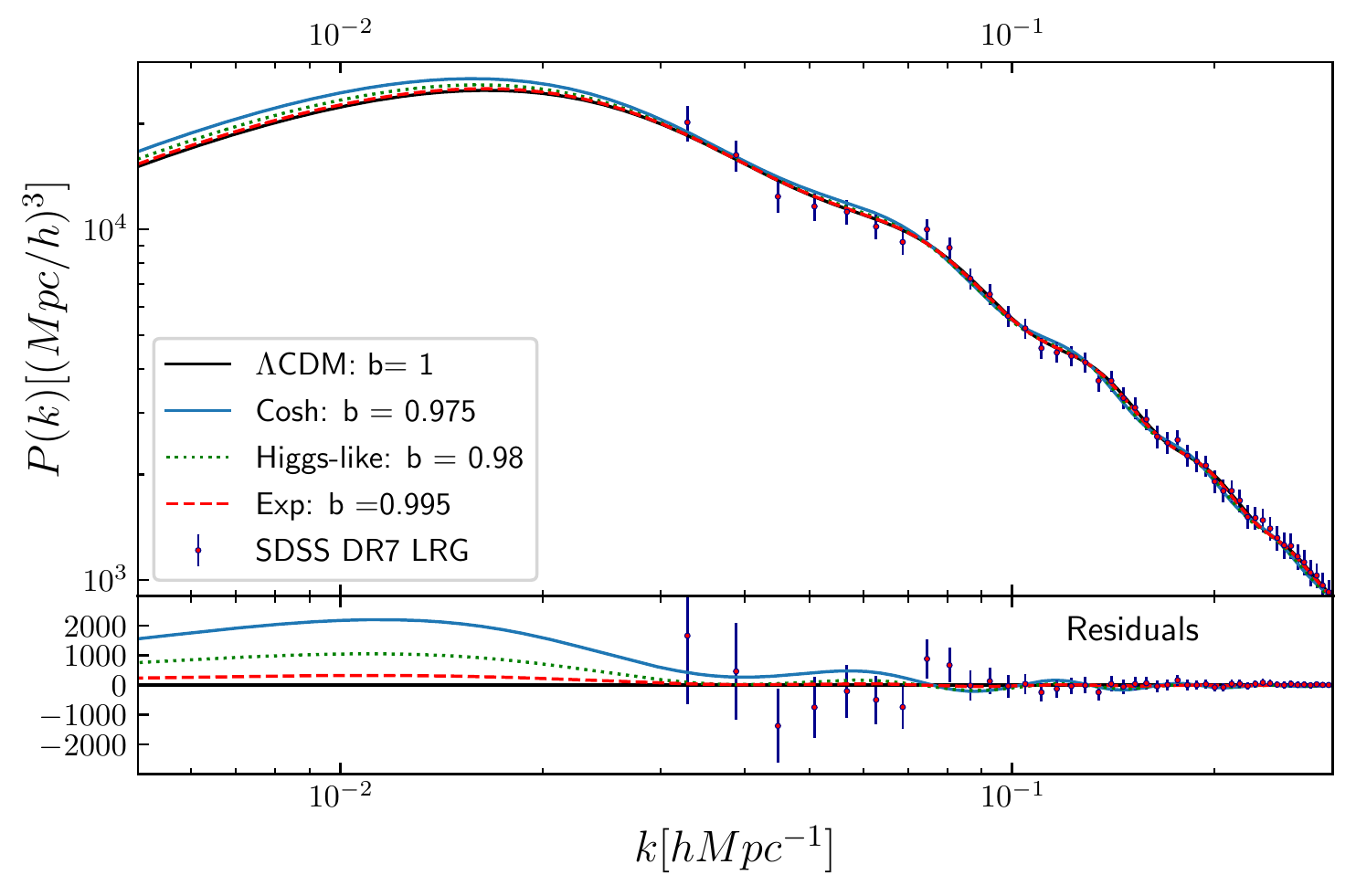}
\end{center}
\vspace{-7mm}
\caption{The linear MPS $P(k)$ for the models of Fig. \ref{Fig:cl} showing excellent fits to the Sloan Digital Sky Survey (SDSS) data release 7 (DR7) luminous red galaxies (LRG)~\cite{Reid:2009xm}. We also include a bias parameter $b$.
Note that the (derived) Hubble constant for each model is different.}
\vspace{-6mm}
\label{Fig:Pk}
\end{figure}

In Figs.~\ref{Fig:cl} and \ref{Fig:Pk} we show the CMB and MPS in the case of a ``Cosh'', an ``Exp'' and a 
``Higgs-like'' function $\Kcal(\Qcal) = \frac{\Kcal_2}{4\Qcal_0^2} \left(\Qcal^2 - \Qcal_0^2\right)^2$,
computed numerically by evolving the  FLRW background and linearized equations using our own Boltzmann code~\cite{Kaplinghat:2002mh},
which is in excellent agreement with other codes, see \cite{Bellini:2017avd} for a comparison. 
We have used adiabatic initial conditions~\cite{SkordisIlicZlosnikInPreparation} and 
a standard initial power spectrum $P_0 = A_s k^{n_s}$ with amplitude $A_s$ and spectral index $n_s$. The MPS has an additional bias parameter $b$.
We used {\footnotesize RECFAST} version 1.5 for modeling recombination and
have boosted $k$ sampling, time sampling and $\ell$ sampling accuracy for ensuring robust results.
The detailed cosmology and the dependence of the spectra on the parameters
 will be investigated elsewhere~\cite{SkordisIlicZlosnikInPreparation}.
For a wide range of parameters, this relativistic MOND theory is
consistent with the CMB measurements from Planck. This happens because $\cad^2$ and $w$ are small enough so that $\Pi \rightarrow 0$
and we get dustlike evolution as
$\dot{\delta} = 3  \dot{\Phi} -  \frac{k^2}{a^2} \theta$ and $\dot{\theta} = \Psi$,  while the vector field decouples.

\paragraph{Stability and waves. --}
Now, we consider stability of the theory on Minkowski spacetime.
We expand $g^{\mu\nu} = \eta^{\mu\nu} - h^{\mu\nu}$, split $\Ah_\mu =(-1+\frac{1}{2} h^{00}, \Ah_i)$ 
 and let $\nabla_\mu\phi =  (\Qcal_0 + \dot{\varphi} , \grad\varphi)$ with $h^{\mu\nu}$, $\Ah_i$ and $\varphi$ being small perturbations.
Expanding  \eqref{NT_A_action} to second order gives 
\begin{widetext}
\begin{align}
S =&  \int d^4x  \bigg\{ 
 - \frac{1}{2} \nablab_\mu h \nablab_\nu h^{\mu\nu}
+ \frac{1}{4}  \nablab_\rho h \nablab^\rho h
+ \frac{1}{2} \nablab_{\mu} h^{\mu\rho}  \nablab_\nu h^\nu_{\;\;\rho}
- \frac{1}{4} \nablab^\rho h^{\mu\nu}  \nablab_\rho  h_{\mu\nu}
  \KB |\dot{\vec{\Ah}} - \frac{1}{2} \grad h^{00}|^2
- 2\KB \grad_{[i} \Ah_{j]} \grad^{[i} \Ah^{j]}
\nonumber
\\
&
+  \left(2 - \KB\right)\left[  2(  \dot{\vec{\Ah}}  -  \frac{1}{2}  \grad h^{00} ) \cdot ( \grad \varphi + \Qcal_0 \vec{\Ah} ) 
- ( 1 + \lambdas )  |\grad\varphi + \Qcal_0 \vec{\Ah}|^2
\right]
+2 \Kcal_2  \left|\dot{\varphi} +  \frac{1}{2}\Qcal_0 h^{00} \right|^2
+  \frac{1}{\Mpt^2} T_{\mu\nu} h^{\mu\nu}
 \bigg\}
\label{Minkowski_action}
\end{align}
\end{widetext}
where we have used the desired late Universe limit for which $\partial^2\bar{\Fcal}/\partial\Qcal^2 \rightarrow -2 d^2\Kcal/d\Qcal^2 = - 4 \Kcal_2 $
and $\partial\Fcal/\partial\Qcal =  \bar{\Fcal} = 0$.  We set $\partial\Fcal/\partial\Ycal  = (2-\KB)\lambdas$ as a free parameter which is
zero in the MOND limit  but nonzero in the GR limit when reached by tracking.  Inspecting \eqref{Minkowski_action}, the tensor mode action is as in GR as expected. 

For vector modes we choose the gauge $h^{00}=0$,  $h^{0i} =W^i$ and  $h^{ij} =0$  while $\Ah_i = \beta_i$ and $\varphi=0$
where $W^i$ and $\beta^i$ are transverse. Setting all modes $\propto e^{i(-\omega t + \vec{k}\cdot\vec{x})}$, 
the dispersion relation for $\beta^i$ is $\omega^2 = k^2 + \Mcal^2$
where their  mass is $ \mass^2 = \frac{(2-\KB)(1 + \lambdas )  \Qcal_0^2}{\KB}$, hence, they  are healthy if $0<\KB<2$ and $\lambdas>-1$.
They decouple from $T_{\mu\nu}$ and are not expected
to be generated to leading order by compact objects.

Considering scalar modes in the Newtonian gauge we set $h^{00} = -2\Psi$, $h^{0i} = 0$ and  $h^{ij} = -2\Phi \gamma^{ij}$ while  $\Ah_i = \grad_i\alpha$
and find the dispersion relations $\omega^2=0$ and 
$\omega^2 = \frac{(2-\KB)}{\Kcal_2 \KB}  (1 + \frac{1}{2} \KB \lambdas) k^2 + \mass^2 $.
Thus, we require that  $\Kcal_2>0$ in addition to the vector stability conditions.
Only two normal modes exist implying the presence of constraints. 
These are revealed through  
a Hamiltonian analysis which also shows that these conditions lead to a positive Hamiltonian~\cite{SkordisZlosnikInPreparationMink,BatakiSkordisZlosnikInPreparation} 
for the $\omega\ne 0$ modes.
The $\omega=0$ case leads to a constant mode with zero Hamiltonian but, also, to a mode varying linearly with $t$. The Hamiltonian for the latter  is positive for momenta
larger than $\sim\mu$ and otherwise negative, also requiring that $\lambdas>0$.
Such instabilities are likely akin to Jeans-type instabilities and do not cause quantum vacuum instability at low momenta~\cite{GumrukcuogluMukohyamaSotiriou2016}.

\paragraph{Discussion. --}
MOND has enjoyed success in fitting galactic rotation curves~\cite{Milgrom1988,Kent1988,Begeman:1991iy,Sanders:1996ua,Sanders:2018mnk} and reproducing the baryonic Tully-Fisher relation~\cite{McGaugh:2000sr,McGaugh:2005qe,LelliEtAl2019}. 
The radial acceleration relation (RAR)~\cite{McGaughEtAl2016} finds a comfortable interpretation within MOND.
Studies of MOND with galaxy clusters~\cite{Sanders1999,Sanders:2002ue,Pointecouteau:2005mr,Famaey:2011kh,Ettori:2018tus,Tian:2020qjd} report that either $a_0$ is larger in clusters and/or an additional dark component is necessary even when the MOND prescription is used.
These studies, however, use the classic modified-inertia MOND while the theory presented here has additional features warranting its separate testing with clusters.
We note that a RAR for clusters was reported~\cite{Tian:2020qjd}, similar to the galaxy one  albeit with $a_0$ a factor of 10 higher. 
MOND has been tested with dwarf spheroidal galaxies where discrepancies for some~\cite{GerhardSpergel1992} were 
later dismissed with improved data~\cite{Milgrom:1995hz,Angus:2008vs,Serra:2009tj,Diez-Tejedor:2016fdn,AlexanderEtal2017}. There,
 good agreement was reported, except for Draco and Carina where the fits are quite poor~\cite{GerhardSpergel1992,Serra:2009tj,AlexanderEtal2017,Read:2018fxs}.
It is argued~\cite{Angus:2008vs} that those two might be systems not in equilibrium.
The global stability of M33 has been tested~\cite{Banik:2020kwt} with positive results while wide-binary data do not yet yield a decisive test~\cite{Pittordis:2019kxq}.

We have shown how the cosmological regime of this theory reproduces the CMB and MPS power spectra on linear scales and that MOND-like behavior emerges in the quasistatic approximation. 
The latter is expected to hold for virialized objects, however, how such objects emerge from the underlying density field, i.e. how the two regimes connect, is
an open problem. This will happen at a scale which is expected to depend on $a_0$, $\mu$ and $\Qcal_0$ and quite likely the nonlinear $\sim \nabla (\nabla \phi)^2/a_0$ term coming from $\Fcal$ will play a role.
It is reasonable to expect that on mildly nonlinear scales, the quasistatic regime is not yet reached.

We remark that $A_\mu$ also contains a pure vector mode perturbation which is expected to behave similarly as 
in the Einstein-{\AE}ther theory~\cite{Dirac1962,JacobsonMattingly2000}. This may lead to imprints on the $B$-mode CMB polarization signal~\cite{Nakashima:2011fu}.

Setting $\Mpt^2 = 1/(8\pi \Gt)$ and canonically normalizing as $\phit =  \sqrt{2\Kcal_2} \Mpt \phi$ in \eqref{Kcal_expansion},
 the FLRW action \eqref{sculpted_FRW_action} becomes
\begin{equation}
 S = \int d^4x N a^3 \bigg[ -3\Mpt^2 \frac{H^2}{N^2} + \frac{1}{2} \left(   \frac{\dot{\tilde{\phi}}}{N}  - \Lambda_c^2\right)^2 + \ldots \bigg]
\end{equation}
where $\Lambda_c^2 =  \Mpt \sqrt{2\Kcal_2}\Qcal_0$. 
Considering the MOND limit  in \eqref{NT_A_action} gives
 $\Mpt^2 \Fcal /2 \rightarrow |\grad\tilde{\phi}|^3 / \Lambda_0^2$ where $\Lambda_0^2 = 12 \left[\Kcal_2(1+ 1/\lambdas)/(2-\KB)\right]^{3/2}   M_p a_0$.
This scale is indicative of the energy scale above which quantum corrections  may be important and below which we can trust the classical theory.
 Since $a_0 \sim H_0/6$ then $\Lambda_0  \gtrsim   \meV \sim  (0.1 \mm)^{-1}$.
Newton's $r^{-2}$ law has been tested down to $\sim 52\mum$~\cite{Lee:2020zjt}  and the curves in  Figs.\ref{Fig:cl} and \ref{Fig:Pk} have $\Lambda_0^{-1} \lesssim 100 \nm$.

Absence of ghosts to quadratic order signifies a healthy theory that could arise as a limit
of a more fundamental theory.  We do not have such a theory at present but we discuss a case that may bring us closer.
The vector in \eqref{NT_A_action} does not seem to obey gauge invariance but in the quadratic action \eqref{Minkowski_action} it does so through mixing with diffeomorphisms of $h_{\mu\nu}$.
This is not an accident.  Let us normalize via $\AC_\mu = \MC \Ah_\mu$ for some scale $\MC$ and insert the term $ -\frac{1}{4}\frac{\Mpt^4}{\MC^4}\lambda^2$.
Varying with $\lambda$ and using the constraint to eliminate $\lambda$ from the action, perform a St\"{u}ckelberg transformation $\AC_\mu \rightarrow \AC_\mu + \nabla_\mu \xi/\MC$
and define the covariant derivative acting on ``angular field'' $\xi$ as $\Dcal_\mu\xi = \nabla_\mu \xi/\MC +  \AC_\mu$. The action turns to
 $S  = S_{EH} +  \int d^4x \sqrt{-\metM} \bigg\{  - \frac{1}{4\ggc^2}  \FC_{\mu\nu}  \FC^{\mu\nu}  
+  \frac{1}{4}  (\Dcal^\mu \xi \Dcal_\mu \xi + \MC^2)^2  \bigg\}
$
plus $\phi$-dependent terms, 
where $\FC_{\mu\nu} = \nabla_\mu \AC_\nu - \nabla_\nu \AC_\mu$, $\ggc^2 = \frac{\MC^2}{\KB \Mpt^2}$. 
The resulting action is that of the gauged ghost condensate (GGC)~\cite{Cheng:2006us} or bumblebee field~\cite{Kostelecky:1988zi,Kostelecky:1989jw} which has been proposed as a healthy gauge-invariant theory 
of spontaneous Lorentz violation.  The Einstein-{\AE}ther theory, part of \eqref{NT_A_action}, is the (healthy) decoupling limit of GGC by taking $\MC\rightarrow \infty$
 if $0<\KB<2$ (in our notation)~\cite{Cheng:2006us}.  It is argued~\cite{Cheng:2006us} that $\MC$ can be as high as $10^{12}\GeV$. 

Given that $\phi$ is shift symmetric it is natural to charge it under this symmetry similar to $\xi$  letting
$\Dcal_\mu \phi = \nabla_\mu\phi/\MC +  \AC_\mu$. Interestingly, we may identify $\Qcal - \Qcal_0 \rightarrow \Dcal^\mu\xi \Dcal_\mu \phi$ while
the term $\Jh^\mu \nabla_\mu \phi \rightarrow \FC^{\mu\nu} \Dcal_\mu \xi  \Dcal_\nu \phi$, both multiplied by appropriate constants.
The terms involving $\Ycal$ may be constructed using $\left(g^{\mu\nu} + \Dcal^\mu \xi \Dcal^\nu \xi/\MC^4\right)\Dcal_\mu\phi \Dcal_\nu \phi$.
Although extending our work as such does not explain the MOND term $\Ycal^{3/2}$, it may provide promising directions for further improvements.

\begin{acknowledgments}
{\it Acknowledgments} We thank C. Burrage, P. Creminelli, S. Ilic, E. Kiritsis, M. Kopp, M. Milgrom, A. Padilla, R. Sanders and I. Sawicki for discussions. 
The research leading to these results has received funding from the European Research Council under the European Union's 
Seventh Framework Programme (FP7/2007-2013) / ERC Grant Agreement No. 617656 ``Theories
 and Models of the Dark Sector: Dark Matter, Dark Energy and Gravity''  and from the European Structural and Investment Funds
and the Czech Ministry of Education, Youth and Sports (MSMT) (Project CoGraDS - CZ.02.1.01/0.0/0.0/15003/0000437).   
\end{acknowledgments}

\bibliographystyle{apsrev}
\bibliography{references}

\end{document}